\documentclass[11pt,showpacs,preprintnumbers,amsmath,amssymb,prd,nofootinbib,superscriptaddress]{revtex4-2}
\usepackage{tikz}
\usetikzlibrary{arrows.meta}
\usepackage{amsmath}
\usepackage{amssymb}
\usepackage{amsbsy}
\usepackage[a4paper, margin=1.8cm]{geometry}
\usepackage{amsfonts}
\usepackage{amsopn}
\usepackage{amstext}
\usepackage{graphicx}
\usepackage{amssymb}
\usepackage{amsfonts}
\usepackage{amsmath}
\usepackage{braket}
\usepackage{graphicx,subcaption}
\usepackage[english]{babel}
\usepackage{color}
\usepackage{xcolor}
\usepackage{slashed}
\usepackage{esint}
\usepackage[dvips]{epsfig}
\usepackage[dvips]{graphicx}
\usepackage{float}
\usepackage{units}
\usepackage{textcomp}
\usepackage{placeins}
\usepackage{hyperref}             
\hypersetup{
    colorlinks=true,              
    breaklinks=true,              
    citecolor=blue,               
    linkcolor=[rgb]{0,0.5,0.9},   
    urlcolor=red,                 
    filecolor=green               
}

\usepackage{hyperref}
\usepackage{slashed}

\newcommand{\ie}{\begin{equation}}
\newcommand{\fe}{\end{equation}}
 \newcommand{\bq}{\begin{equation}}
 \newcommand{\eq}{\end{equation}}
 \newcommand{\bqn}{\begin{eqnarray}}
 \newcommand{\eqn}{\end{eqnarray}}

 \newcommand{\orcid}[1]{\href{https://orcid.org/#1}{\includegraphics[width=10pt]{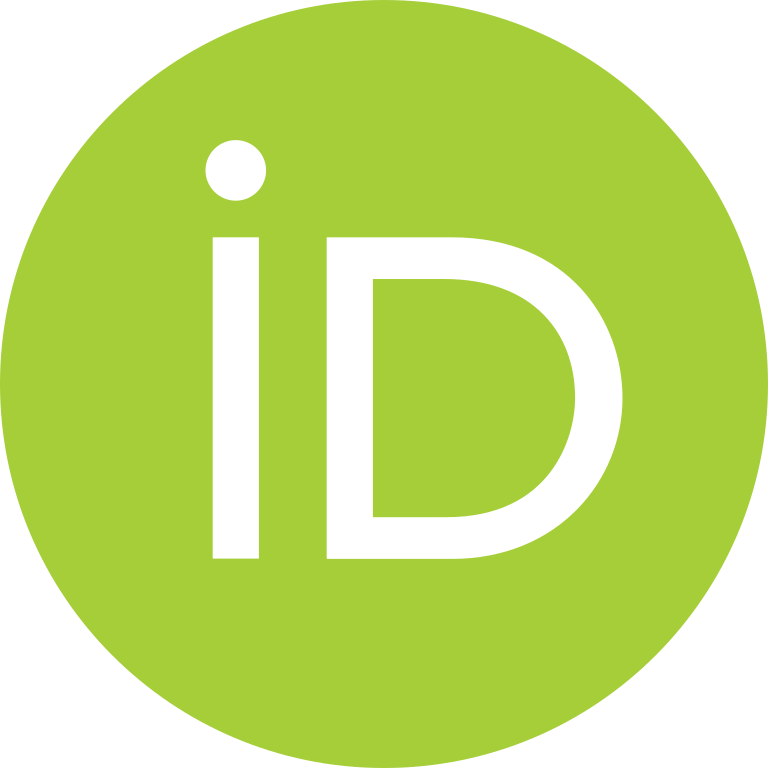}}}

\begin{document}

    \title{Geometric and Statistical Thermo Field Dynamics in de Sitter Spacetime}
	
	\author{D. S. Cabral  \orcid{0000-0002-7086-5582}}
	\email{danielcabral@fisica.ufmt.br}
	\affiliation{Programa de P\'{o}s-Gradua\c{c}\~{a}o em F\'{\i}sica, Instituto de F\'{\i}sica,\\ 
		Universidade Federal de Mato Grosso, Cuiab\'{a}, Brasil}
	
	\author{L. A. S. Evangelista \orcid{0009-0002-3136-2234}}
	\email{lucassouza@fisica.ufmt.br}
	\affiliation{Programa de P\'{o}s-Gradua\c{c}\~{a}o em F\'{\i}sica, Instituto de F\'{\i}sica,\\ 
		Universidade Federal de Mato Grosso, Cuiab\'{a}, Brasil}
	
	\author{J. C. R. de Souza \orcid{0000-0002-7684-9540}}
	\email{jean.carlos@fisica.ufmt.br}
	\affiliation{Programa de P\'{o}s-Gradua\c{c}\~{a}o em F\'{\i}sica, Instituto de F\'{\i}sica,\\ 
		Universidade Federal de Mato Grosso, Cuiab\'{a}, Brasil}
	
	\author{A. F. Santos \orcid{0000-0002-2505-5273}}
	\email{alesandroferreira@fisica.ufmt.br}
	\affiliation{Programa de P\'{o}s-Gradua\c{c}\~{a}o em F\'{\i}sica, Instituto de F\'{\i}sica,\\ 
		Universidade Federal de Mato Grosso, Cuiab\'{a}, Brasil}

\begin{abstract}

The dynamics of a massive scalar field non-minimally coupled to gravity in an expanding de Sitter universe are investigated. It is shown that a comoving observer identifies the Bunch--Davies state as the vacuum, whereas a static observer perceives the same state as a thermal bath at the Gibbons--Hawking temperature. Motivated by this observer dependence, a thermal formulation based on Thermo Field Dynamics is developed by combining the geometric doubling associated with the cosmological horizon with the statistical doubling induced by a intrinsec thermal bath. The resulting construction reveals that the doubling procedure is not merely a mathematical artifact, but rather a manifestation of the global causal structure of spacetime together with finite-temperature effects. The temporal evolution of the Bogoliubov angle is analyzed and the corresponding particle number densities are evaluated in both comoving and static frames. In the radiation limit, the comoving number density remains conserved, providing a thermodynamic evolution consistent with that of the Cosmic Microwave Background, whereas in the static frame finite-temperature effects sti\-mulate Parker particle creation. For massive and non-minimally coupled fields, the interplay between geometric and statistical temperatures gives rise to a characteristic thermal scale and a nontrivial dependence on the initial conditions. These results provide a unified framework for describing quantum fields in de Sitter spacetime in the presence of both apparent horizon-induced and intrinsec thermal effects.
		
\end{abstract}

\maketitle

\section{Introduction}

The interplay between quantum field theory and gravitation emerged through Hawking's seminal analysis of quantum fields in black-hole spacetimes \cite{hawking1974black,hawking1975particle}. It was shown that event horizons radiate thermally, with a temperature determined by their surface gravity, implying that black holes possess genuine thermodynamic properties and can gradually evaporate. This framework was later generalized to cosmological spacetimes with a positive cosmological constant \cite{gibbons1977cosmological}, where the presence of a cosmological horizon gives rise to the analogous Gibbons--Hawking temperature. Together, these discoveries established a profound relationship between quantum field theory, thermodynamics, and the geometrical and causal structure of spacetime. In such spacetimes, the cosmological horizon acquires thermodynamic properties analogous to those of black-hole horizons, with its area being proportional to an entropy associated with the information inaccessible beyond the horizon. As a consequence, the notions of particles and thermal radiation become intrinsically observer dependent. This remarkable feature was elucidated by Unruh \cite{unruh1976notes}, who demonstrated that particle content depends on the observer's state of motion: while a freely falling observer crossing the horizon does not detect thermal radiation locally, a stationary observer in a gravitational field, or equivalently a uniformly accelerated observer, perceives a thermal bath of particles.

Although these developments provide a consistent geometrical and thermodynamical description of black-holes and cosmological spacetimes, they do not fully elucidate the algebraic structure relating the physical descriptions associated with different observers. In this context, Israel \cite{israel1976thermo} established a geometrical correspondence between the two causally disconnected regions separated by the horizon, revealing a structure closely related to the Thermo Field Dynamics (TFD) formalism introduced by Takahashi and Umezawa \cite{Umezawa1,Umezawa2}. In TFD, thermal expectation values are represented as vacuum expectation values in an enlarged Hilbert space, $\mathcal{H}_T$, obtained by doubling the physical degrees of freedom according to $\mathcal{H}_T=\mathcal{H}\otimes\widetilde{\mathcal{H}}$, where $\mathcal{H}$ denotes the physical Hilbert space and $\widetilde{\mathcal{H}}$ its auxiliary tilde counterpart. Within this framework, Israel showed that the Kruskal vacuum, which represents the global vacuum state, can be interpreted as an entangled state composed of correlated particle pairs distributed across both sides of the horizon and related through Bogoliubov transformations. However, his analysis was restricted to the geometric thermal properties induced by the horizon and did not incorporate statistical thermal effects associated with the presence of a genuine thermal bath.

The distinction between geometric thermality and genuine statistical thermal effects becomes particularly relevant in the context of the inflationary universe. During the inflationary epoch, quantum fluctuations were stretched beyond the de Sitter horizon, providing the primordial seeds for the anisotropies and large-scale structures observed in the Cosmic Microwave Background (CMB) \cite{starobinsky1982dynamics,hawking1982development}. Although the standard inflationary scenario assumes that these fluctuations originate from a pure Bunch-Davies vacuum, alternative mechanisms, such as warm inflation \cite{berera1995warm} or particle production during the inflationary stage, suggest that quantum fields may instead evolve from a thermalized Bunch-Davies state embedded in a statistical plasma. In such a scenario, the primordial perturbation spectrum may contain signatures arising from the interplay between the geometric temperature associated with the cosmological horizon and the statistical temperature of the surrounding thermal bath. Understanding how these two distinct sources of thermality coexist and manifest themselves therefore becomes essential for a complete description of quantum fields in the early universe.

In this work, Israel's construction is extended to de Sitter spacetime by incorporating not only the geometric doubling associated with the cosmological horizon, but also the statistical thermal structure inherent to TFD. To this end, Right and Left sectors are introduced, corresponding to the observable and causally disconnected regions of de Sitter spacetime, together with a thermal copy of each sector describing a physical thermal bath. The resulting Hilbert space takes the form
$\mathcal{H}_\beta=\mathcal{H}_R\otimes\widetilde{\mathcal{H}}_R\otimes\mathcal{H}_L\otimes\widetilde{\mathcal{H}}_L$,
thereby providing a unified framework in which causal and thermal degrees of freedom are treated on equal footing. Within this formalism, the particle creation process is investigated in the presence of a genuine thermal bath, and possible implications for the primordial spectrum and the Cosmic Microwave Background are discussed. For general discussions of the structural aspects of TFD in quantum field theory in Minkowski spacetime and its applications in various contexts, see Refs.~\cite{Book,Khanna1,Khanna2,ref1,ref2,ref3}.

This paper is organized as follows. In Section~\ref{2}, the dynamics of a massive scalar field non-minimally coupled to gravity is investigated in a de Sitter background. In Section~\ref{3}, the geometric formulation of Thermo Field Dynamics is developed. The Kruskal coordinates are introduced, the corresponding Kruskal diagram is analyzed, and the four sectors of the maximally extended spacetime are discussed. It is shown that a comoving observer identifies the Bunch--Davies state as the vacuum, whereas a static observer perceives the same state as a thermal bath at the Gibbons--Hawking temperature. In Section~\ref{4}, finite-temperature effects are incorporated into the global vacuum state, allowing both the geometric and genuine thermal contributions to be examined. The temporal evolution of the Bogoliubov angle is studied and the corresponding comoving number density is evaluated. Finally, concluding remarks are presented in Section~\ref{5}.

\section{Field Dynamics in de Sitter Spacetime}\label{2}

In this section, the dynamics of a massive scalar field non-minimally coupled to gravity are investigated in a de Sitter spacetime background. For this purpose, the Lagrangian density of the field is taken as
\begin{equation}
    \mathcal{L}=\sqrt{-g}\,\left[\frac{1}{2}\nabla_{\mu}\phi\nabla^{\mu}\phi-\frac{1}{2}M^2\phi^2-f(\phi)\mathcal{R}\right],\label{eq12}
\end{equation}
where $g$ denotes the determinant of the metric tensor, $\nabla_\mu$ is the covariant derivative, $M$ is the scalar field mass, $\mathcal{R}$ is the Ricci scalar, and $f(\phi)$ represents the non-minimal coupling function. By assuming the form $f(\phi)=\frac{1}{2}\xi\phi^2$, with $\xi$ being the non-minimal coupling parameter, the corresponding equation of motion is obtained as
\begin{eqnarray}
    \square\phi+M^2\phi+\xi \mathcal{R}\phi=0.\label{eq01}
\end{eqnarray}
The spacetime geometry is described by the Friedmann-Lemaître-Robertson-Walker (FLRW) metric,
\begin{equation}
\mathrm{d}s^2 = \mathrm{d}t^2 -a(t)^2\biggl ( \frac{\mathrm{d}r^2}{1-k\,r^2}+ r^2 \, \mathrm{d}\theta^2 +r^2 \, \sin ^2\theta \,\mathrm{d}\phi^2 \biggr ),
\label{eq00}
\end{equation}
where $a(t)$ is the scale factor and $k$ determines the spatial curvature. The metric signature is chosen as $\mathrm{diag}(1,-1,-1,-1)$. In the following, attention is restricted to the spatially flat case ($k=0$) and to a de Sitter background, for which the scale factor is given by $a(t)=e^{Ht}$,
with $H$ denoting the Hubble parameter.

Expressed in terms of the cosmic time $t$, and assuming the mode decomposition $\phi(t,\vec{x}) \propto e^{-i\vec{p}\cdot\vec{x}}\,q(t)$, Eq.~\eqref{eq01} reduces to
\begin{eqnarray}
    \ddot{q}(t)+3H\dot{q}(t)+\omega^2q(t)=0,\quad\quad\text{with}\quad\quad \omega^2=|\vec{p}\;|^2e^{-2Ht}+M^2+12\xi H^2.\label{eq16}
\end{eqnarray}
Here, $\omega(t)$ denotes the instantaneous energy of the particle, while the Ricci scalar is given by $\mathcal{R}=12H^2$. Introducing the conformal time $\eta$, defined by $\eta^{-1}=-Ha(\eta)$, and the rescaled auxiliary function $y(\eta)=a(\eta)q(\eta)$, Eq.~\eqref{eq16} reduces to
\begin{eqnarray}
    \frac{d^2y(\eta)}{d\eta^2}+\left[\vec{p\;}^2+\frac{1}{\eta^2}\left(\frac{M^2}{H^2}+12\xi-2\right)\right]y(\eta)=0,\
\end{eqnarray}
where the quantity enclosed in brackets is identified as the squared effective cosmological frequency. The solution to this equation is given by
\begin{eqnarray}
\phi(\eta,r,\Omega)=\sum_{\ell,m}\int_0^\infty dp\left(N_pH\eta\right)\left[\sqrt{\eta}H_\nu^{(2)}(p\eta)j_\ell(pr)Y_{\ell m}(\Omega)b_{\ell m}(p)+\text{h.c.}\right],\label{eq07}
\end{eqnarray}
where $\Omega$ denotes the solid angle, $N_p$ is a normalization constant, and $\nu^2=\left(\frac{9}{4}-\frac{M^2}{H^2}-12\xi\right)$ determines the order of the temporal modes. Furthermore, $H_\nu^{(2)}$ is the Hankel function of the second kind, while $j_\ell$ and $Y_{\ell m}$ denote the spherical Bessel functions and spherical harmonics, respectively. The operator $b_{\ell m}(p)$ corresponds to the annihilation operator. In this representation, the creation and annihilation operators satisfy
\begin{eqnarray}
    b_{\ell m}(p)\ket{0_\text{BD}}=0,\quad\quad \text{with} \quad\quad \left[b_{\ell m}(p),b_{\ell^\prime m^\prime}^\dagger(p^\prime)\right]=\delta(p-p^\prime)\delta_{\ell,\ell^\prime}\delta_{m,m^\prime},\label{eq10}
\end{eqnarray}
where $\ket{0_\text{BD}}$ denotes the Bunch-Davies vacuum state, namely the de Sitter-invariant vacuum obtained by matching the field modes to positive-frequency Minkowski modes in the early-time limit ($\eta\to-\infty$).

The solution \eqref{eq07} is selected by requiring that, in the short-wavelength limit, where the physical wavelength is much smaller than the Hubble radius, the effects of the cosmological expansion become negligible. In this regime, the temporal modes reduce to positive-frequency plane waves with respect to the conformal time, namely,
\begin{eqnarray}
\lambda\sim \frac{2\pi}{p|\eta|}\ll H^{-1}\longrightarrow p|\eta|\gg1.
\label{eq18}
\end{eqnarray}
This condition uniquely determines the Bunch-Davies vacuum and ensures that the field modes asymptotically coincide with those of Minkowski spacetime in the ultraviolet limit \cite{bunch1978quantum}.

The preceding discussion and the solution given by Eq.~\eqref{eq07} describe a field quantized in the comoving frame associated with the expanding de Sitter universe. The corresponding causal structure is determined by the null condition $\mathrm{d}s^2=0$. In order to describe the field from the viewpoint of another observer, a suitable coordinate transformation is required. As will be shown below, this transformation leads naturally to the emergence of the fundamental ingredients of the Thermo Field Dynamics formalism, which is presented in the next section.

\section{Geometric Thermo Field Dynamics}\label{3}

This section is devoted to the geometrical representation of TFD, following the approach presented in Ref.~\cite{israel1976thermo}. Particular attention is given to the reinterpretation of these results within the conventional TFD framework, a perspective that has not been extensively explored in the context of that work.

In order to describe the field from the perspective of static observers, the coordinates $T$ and $R$ are introduced according to
\begin{eqnarray}
	R=e^{Ht}r,\quad\quad\text{and}\quad\quad  dt=dT-\frac{HR}{1-H^2R^2}dR.
\end{eqnarray}
In terms of these coordinates, the metric \eqref{eq00} can be written as
\begin{eqnarray}
\mathrm{d}s^2 = (1-H^2R^2)\mathrm{d}T^2 -\frac{\mathrm{d}R^2}{(1-H^2R^2)}- R^2 \, \mathrm{d}\Omega^2.
\end{eqnarray}
This line element describes the spacetime as perceived by a static observer with respect to the time coordinate $T$ and exhibits a cosmological horizon located at $R=H^{-1}$.

In this coordinate system, the equation of motion \eqref{eq01} admits the solution
\begin{eqnarray}
	\phi(T,R,\Omega)=\sum_{\ell,m}\int_0^\infty d\omega\frac{N_\omega}{R}\left[e^{-i\omega T}\psi_{\omega\ell}(R)Y_{\ell m}(\Omega)c_{\ell m}(\omega)+\text{h.c.}\right],
\end{eqnarray}
where the creation and annihilation operators satisfy commutation relations analogous to those in Eq.~\eqref{eq10}, but are defined with respect to the vacuum state perceived by static observers, denoted by $\ket{0_\text{static}}$.

Furthermore, the radial modes $\psi_{\omega\ell}$ satisfy the effective Schrödinger-like equation
\begin{eqnarray}
\frac{d^2\psi_{\omega\ell}}{d\rho^2}+\left[\omega^2-V_\text{eff}(R)\right]\psi_{\omega\ell}=0.
\end{eqnarray}
Here, $\rho$ denotes the tortoise coordinate, defined by
\begin{eqnarray}
\rho=\frac{1}{2H}\ln\left(\frac{1+HR}{1-HR}\right),
\end{eqnarray}
while the effective potential is given by
\begin{eqnarray}
V_\text{eff}(R)=(1-H^2R^2)\left[\frac{\ell(\ell+1)}{R^2}+M^2+\xi \mathcal{R}-2H^2\right].
\end{eqnarray}

In the near-horizon limit, $V_\text{eff}\approx 0$, the solution of the effective Schrödinger-like equation takes the form
\begin{eqnarray}
	\psi_{\omega\ell}(\rho)=\alpha e^{-i\omega \rho}+\beta e^{i\omega\rho}.
\end{eqnarray}
By introducing the null coordinates $u=T-\rho$ and $v=T+\rho$, the complete solution is expressed as
\begin{eqnarray}
	\phi(u,v,\Omega)=\sum_{\ell,m}\int_0^\infty d\omega\frac{N_\omega}{R}\left[e^{-i\omega u}Y_{\ell m}(\Omega)c^{+}_{\ell m}(\omega)+ e^{-i\omega v}Y_{\ell m}(\Omega)c^{-}_{\ell m}(\omega)+\text{h.c.}\right].\label{eq05}
\end{eqnarray}
Here, the coefficients $\alpha$ and $\beta$ (or $+$ and $-$) correspond to ingoing and outgoing propagation modes, respectively.

The Kruskal coordinates are defined as
\begin{eqnarray}
	U=-e^{-Hu}\quad\quad\text{and}\quad\quad V=e^{Hv},
\end{eqnarray}
which implies $U<0$ and $V>0$. Substituting these into the field solution \eqref{eq05} yields
\begin{align}
	\phi_\text{static}(U,V,\Omega)=\sum_{\ell,m}\int_0^\infty d\omega\frac{N_\omega}{R}&\biggl[\left(-U\right)^{i\omega/H}Y_{\ell m}(\Omega)c^{+}_{\ell m}(\omega)\nonumber\\&+\left(V\right)^{-i\omega/H}Y_{\ell m}(\Omega)c^{-}_{\ell m}(\omega)+\text{h.c.}\biggr].\label{eq09}
\end{align}
This expression exhibits a non-analytic behavior at $U=0$ or $V=0$, corresponding to the cosmological horizons, and is therefore restricted to a particular region of spacetime.

Expressed in Kruskal coordinates, the metric can be written as
\begin{eqnarray}
	\mathrm{d}s^2=\frac{4dUdV}{H^2(1-UV)^2}-R^2d\Omega^2,
\end{eqnarray}
where the radial coordinate $R$ is implicitly given by
\begin{eqnarray}
	R=-\frac{1}{H}\frac{1+UV}{1-UV}.
\end{eqnarray}

\begin{figure}[!htb]
    \centering
    \includegraphics[width=0.5\linewidth]{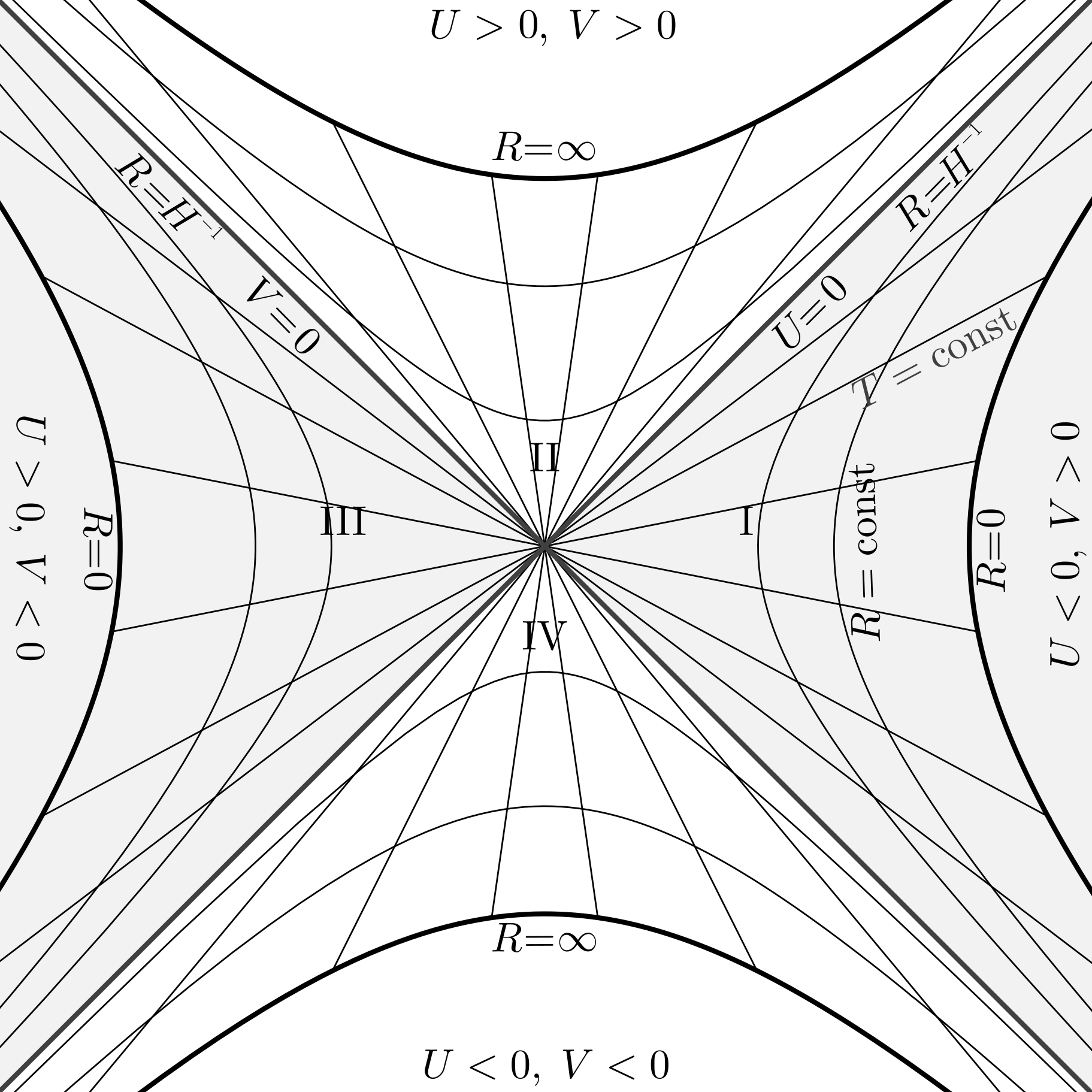}
    \caption{The Kruskal diagram of the maximally extended space-time, split into four quadrants: region I at the right; region II beyond the event horizon of the former; region III at the left, a mirrored copy of the right quadrant; and region IV, containing the space-time points beyond the past event horizon.}
    \label{kruskal}
\end{figure}

As shown in Figure~\ref{kruskal}, the Kruskal diagram of the maximally extended de Sitter spacetime is displayed. The four regions represented in the diagram are discussed below. Region I, also referred to as the right region ($U<0$ and $V>0$), corresponds to the static patch causally accessible to an observer following a timelike trajectory confined within the cosmological horizon. In this region, the causal structure preserves the usual interpretation of the temporal and radial coordinates, so the observer remains within the causal communication domain associated with the center of the diagram. The signal $UV<0$, precisely characterizes the region inside the causal cone of the original domain. The hyperboles $R=\mathrm{const}$ remain regular in this sector, and the radius associated with the observable region remains finite as long as $UV$ does not approach the critical value that defines the horizon. In physical terms, this is the locally observable domain of spacetime, in which cosmological time $t$ and radial coordinate $r$ admit direct interpretation and in which the observer can still causally probe by the observer.

Region II ($U>0$ and $V>0$) is separated from the previous one by a null surface and corresponds to the maximally analytical extension located beyond the cosmological horizon of the static domain. The fact that $UV>0$ implies that the two null coordinates have the same sign, which, identifies a sector in which the causal structure is reordered in relation to the original domain. Consequently, this region is not accessible by timelike trajectories originating in the previous region without crossing the horizon, which should not be interpreted as a material singularity but as a causal boundary of the observable domain. From a global point of view, the geometry remains regular in this sector; the impossibility of access stems exclusively from causality and not from a geometric divergence. Thus, Kruskal's diagram makes explicit that the cosmological horizon of the static system is a coordinate boundary of the observer and not a physical singularity of the manifold.

Region III, also referred to as the left region ($U>0$ and $V<0$) constitutes the mirror sector of the accessible static region, with the same signature $UV<0$, but with the orientation of the signs $U$ and $V$ inverted. This inversion distinguishes a causally disconnected quadrant from the original observable domain, separated by null horizons and accessible only by analytical extension of the solution. In geometric terms, it is a dual region, or second static copy, that reproduces the local structure of the original region without allowing direct causal communication with it. In physical terms, the structural duplication does not imply the coexistence of interacting realities, but only the presence of a global symmetry of the maximally extended solution. The local observer does not have operative access to this sector, although it exists as an integral part of the global manifold. Thus, this region shows that the observed domain is only a causal cutout of the complete geometry and that the solution admits a causally disconnected mirror extension of region I.

Region IV ($U<0$ and $V<0$) is maximally analytic extension and completes the symmetrical structure of the manifold. Although it also satisfies $UV>0$, as the region II, corresponding to the sector beyond the horizon of the original domain, in this case both signs are negative, which places it in the opposite quadrant to the dual region. The natural interpretation is that of a region situated beyond the horizon of the dual copy itself, playing, in relation to that copy, the same role that the previous region plays in relation to the original universe.

The maximally extended space-time is organized into pairs of sectors related by causal reflection: an accessible static region, a region beyond the horizon, a causally disconnected dual region, and a region beyond the horizon of the dual. This pairing expresses the global symmetry of the analytic extension. In the $(U,V)$ plane, the passage between quadrants is equivalent to traversing null horizons and reclassifying the causal sector of the solution, which makes the global structure of space-time explicit.

The field can be analyzed in both Regions I and III, corresponding to the right and left static patches of de Sitter spacetime. However, the mode expansion expressed in terms of the Kruskal coordinates in Eq.~\eqref{eq09} is restricted to Region I and therefore does not provide a complete description of the field over the entire manifold $\mathcal{M}$. A complete set of modes must be defined on both static sectors of the maximally extended spacetime. To this end, let us return to the action of the scalar field theory
\begin{eqnarray}
	S=\int_\mathcal{M} d^4x\mathcal{L}_\text{total}=\int_{\substack{\text{right}\\ \text{region}}}d^4x_R\mathcal{L}_R(U_R,V_R,\Omega)+\int_{\substack{\text{left}\\ \text{region}}}d^4x_L\mathcal{L}_L(U_L,V_L,\Omega)
\end{eqnarray}
where the first term corresponds to the theory developed thus far, while the second is obtained from the former by interchanging the coordinates $U$ and $V$.

The solution for the dual theory is given by
\begin{align}
	\phi_L(U_L,V_L,\Omega)=\sum_{\ell,m}\int_0^\infty d\omega\frac{N_\omega}{R}&\biggl[\left(-V_L\right)^{i\omega/H}Y_{\ell m}(\Omega){c}^{\;+}_{\ell m L}(\omega)\nonumber\\&+ \left(U_L\right)^{-i\omega/H}Y_{\ell m}(\Omega){c}^{\;-}_{\ell m L}(\omega)+\text{h.c.}\biggr].
\end{align}
Following the analysis depicted in Figure~\ref{kruskal}, we observe that $U_R=-U_L$ and $V_R=-V_L$, such that
\begin{eqnarray}
	S=\int_\mathcal{M} d^4x\mathcal{L}_\text{total}=\int d^4x\left[\mathcal{L}_R(U,V,\Omega)-\mathcal{L}_L(U,V,\Omega)\right]\label{eq13}
\end{eqnarray}
which implies
\begin{align}
	\phi_\text{static}^L(U,V,\Omega)=\sum_{\ell,m}\int_0^\infty d\omega\frac{N_\omega}{R}&\biggl[ \left(V\right)^{i\omega/H}Y_{\ell m}(\Omega)c^{\;+}_{\ell m L}(\omega)\nonumber\\&+ \left(-U\right)^{-i\omega/H}Y_{\ell m}(\Omega)c^{\;-}_{\ell m L}(\omega)+\text{h.c.}\biggr].
\end{align}
In this representation, two sets of quantities are defined: the operators and states of the observable region of the universe, which belong to the Hilbert space $\mathcal{H}_R$, and those of the left (dual) region, which lie in the Hilbert space $\mathcal{H}_L$. Consequently, doubled quantities may be introduced, such as the field operators and their conjugate momenta,
\begin{eqnarray}
	\varphi_\text{static}=\begin{pmatrix}
		\phi_\text{static}^R\\\phi_\text{static}^{L\;\dagger}
	\end{pmatrix},\quad\quad \overline{\varphi}_\text{static}=\begin{pmatrix}
		\phi_\text{static}^{R\;\dagger}&-\phi_\text{static}^L
	\end{pmatrix}.\label{eq11}
\end{eqnarray}

Thus, while the individual solutions $\phi_\text{static}^R$ and $\phi_\text{static}^L$ are restricted to describing the scalar field in the right and left regions, respectively, the doubled field $\varphi_\text{static}$ constitutes a single unified mathematical object capable of covering the entire extended manifold.

The global solution, which describes the entire spacetime and extends the Bunch-Davies (BD) solution presented in Eq.~\eqref{eq07}, includes the static observer as a particular case and must be regular at the horizons. Through analytic continuation, the Kruskal coordinates yield
\begin{eqnarray}
   \left(-U\right)^{i\omega/H}=e^{-\pi\omega/H}|U|^{i\omega/H},\quad\quad  \left(V\right)^{-i\omega/H}=e^{\pi\omega/H}|V|^{-i\omega/H}\label{eq06}
\end{eqnarray}
which reveals a thermal spectrum characterized by the Hawking temperature $T_H=H/(2\pi)$. Consequently, the Kruskal extension of the BD field is given by
\begin{eqnarray}
    \phi_\text{BD}^R(U,V,\Omega)=\sum_{\ell,m,\alpha}\int_0^\infty d\omega \frac{N_\omega}{R}\left[g^{\alpha}_{\ell mR}(\omega)Y_{\ell m}(\Omega)b^{\alpha}_{\ell mR}+\text{h.c.}\right]\label{eq08}
\end{eqnarray}
where $\alpha=\pm1$. A similar expression can be obtained for $\phi_\text{BD}^L$, allowing us to define the total field in this representation as $\varphi_\text{BD}=\begin{pmatrix}\phi_\text{BD}^R&\phi_\text{BD}^{L\;\dagger}
\end{pmatrix}^T$.

Note that $\varphi_\text{BD}=\varphi_\text{static}$ because the underlying physics is invariant under diffeomorphisms. Furthermore, the annihilation operators satisfy
\begin{eqnarray}
    b_{\ell mR}\ket{0_\text{BD}}=0,\quad\quad c_{\ell m R}\ket{0_\text{static}}=0
\end{eqnarray}
and the same applies to the $b_{\ell mL}$ and $c_{\ell mL}$ operators. Here, the vacuum states are factorized as $\ket{0_\text{BD}}=\ket{0^R_\text{BD}}\otimes\ket{0^L_\text{BD}}$ and $\ket{0_\text{static}}=\ket{0^R_\text{static}}\otimes\ket{0^L_\text{static}}$. This implies that
\begin{eqnarray}
    b_{\ell m}\neq c_{\ell m},\quad\quad\text{and consequently}\quad\quad \ket{0_\text{BD}}\neq\ket{0_\text{static}}.
\end{eqnarray}
To obtain a globally defined scalar field on the extended manifold, appropriate correspondence relations between the modes in the different regions must be established. These relations are implemented through Bogoliubov transformations. By rewriting Eq.~\eqref{eq08} and taking into account Eqs.~\eqref{eq06} and \eqref{eq09}, the mode functions can be expressed as
\begin{align}
    g_{\ell mR}^{+}&=\frac{1}{\sqrt{1-e^{-2\pi\omega/H}}}\left[\Theta(-U)(-U)^{i\omega/H}+\Theta(U)e^{-\pi\omega/H}U^{i\omega/H}\right],\nonumber\\ g_{\ell mR}^{-}&=\frac{1}{\sqrt{1-e^{-2\pi\omega/H}}}\left[\Theta(V)(V)^{-i\omega/H}+\Theta(-V)e^{-\pi\omega/H}(-V)^{-i\omega/H}\right],
\end{align}
where $\Theta(x)$ is the Heaviside step function. Analogous expressions are obtained for the modes associated with the left region. These Bogoliubov modes are orthonormalized with respect to the standard Klein-Gordon inner product
\begin{eqnarray}
    (f_1,f_2)=-i\int_\Sigma d^3x\sqrt{\mathfrak{g}_\Sigma}\left(f_1\nabla_\mu f_2^*-f_2^*\nabla_\mu f_1\right)n^\mu,
\end{eqnarray}
where $\Sigma$ denotes a spacelike Cauchy hypersurface, $\sqrt{\mathfrak{g}_\Sigma}$ is the determinant of the induced metric on $\sigma$, while $n^\mu$ is the future-directed unit normal vector. Consequently, the complete solution extends from the actual universe, across the horizon, into the dual universe, and vice versa.

The complete set of modes gives rise to the following Bogoliubov transformations
\begin{align}
	C_{\ell m}^{\alpha}=\mathbb{M}(\beta_H)B_{\ell m}^{\alpha}.
\end{align}
In matrix form, analogously to Eq.~\eqref{eq11}, the Bogoliubov transformations are defined as
\begin{eqnarray}
	C_{\ell m}^{\alpha}=\begin{pmatrix}
		c_{\ell mR}^{\alpha}\\c_{\ell mL}^{\alpha\;\dagger}
	\end{pmatrix};\quad\quad 
	\mathbb{M}(\beta_H)=\begin{pmatrix}
		\cosh{\theta_\omega(\beta_H)} &   \sinh{\theta_\omega(\beta_H)}\\
		\sinh{\theta_\omega(\beta_H)} &   \cosh{\theta_\omega(\beta_H)}
	\end{pmatrix},\label{eq14}
\end{eqnarray}
where $\beta_H=2\pi/H$ and
\begin{eqnarray}
	\sinh{\theta_\omega(\beta)}=\frac{e^{-\beta \omega/2}}{\sqrt{1-e^{-\beta\omega}}},\quad\quad\text{and}\quad\quad \cosh{\theta_\omega(\beta)}=\frac{1}{\sqrt{1-e^{-\beta\omega}}},\label{eq19}
\end{eqnarray}
while the column vector $B_{\ell m}^{\alpha}$ is obtained from $C_{\ell m}^{\alpha}$ by replacing the operators $c_{\ell m}$ with $b_{\ell m}$.

Consequently, the following expression is obtained for the expectation value of the number operator:
\begin{eqnarray}
	\bra{0_\text{BD}}c^{\alpha\;\dagger}_{\ell mR}(\omega)c^{\alpha}_{\ell mR}(\omega)\ket{0_\text{BD}}=\left(e^{2\pi \omega/H}-1\right)^{-1},\label{eq25}
\end{eqnarray}
which characterizes a thermal bath at inverse temperature $\beta_H$. The corresponding temperature,
$T_H=\frac{H}{2\pi},$ is known as the Gibbons--Hawking temperature \cite{gibbons1977cosmological}. The same result is found for the left region. This result reflects the particle production associated with the expanding de Sitter background, commonly known as the Parker effect. Consequently, while a comoving observer identifies the Bunch--Davies state as the vacuum, a static observer perceives the same state as a thermal bath at the Gibbons--Hawking temperature $T_H$ \cite{parker1968particle,gibbons1977cosmological}.

This result indicates that the Bunch-Davies vacuum can be expressed in terms of the static basis as follows
\begin{eqnarray}
	\ket{0_\text{BD}}=\sqrt{1-e^{-2\pi\omega/H}}\sum_{n=0}^\infty e^{-n\pi\omega/H}\ket{n_{\text{static}}^R}\otimes\ket{n_{\text{static}}^L}.\label{eq17}
\end{eqnarray}
These relations can be readily inverted through the inverse of the Bogoliubov matrix \eqref{eq14}, thereby allowing the static vacuum state $\ket{0_\text{static}}$, as well as all physical quantities defined in the static frame, to be expressed in terms of the Bunch--Davies basis. This construction establishes the Thermo Field Dynamics formalism in an expanding universe. It provides a natural framework for defining doubled Lagrangian and Hamiltonian operators, constructing field doublets, introducing dual creation and annihilation operators, and implementing Bogoliubov transformations, in close analogy with the formulation proposed by Israel \cite{israel1976thermo}.

The primary difference between the present geometric construction and the original Thermo Field Dynamics formalism proposed by Takahashi and Umezawa \cite{Umezawa1,Umezawa2} lies in the origin of the temperature. In the present framework, the thermal behavior emerges from the geometric and causal structure of de Sitter spacetime through the relation between the Bunch--Davies and static vacuum states. As a consequence, the Bunch--Davies vacuum is perceived by a static observer as a thermal state at the Gibbons--Hawking temperature. This observer-dependent thermality should not be interpreted as the presence of a physical thermal bath of scalar particles filling spacetime, nor as a universe maintained at an externally prescribed finite temperature.

To incorporate a comoving frame immersed in a genuine thermal bath and thereby describe an expanding universe with finite-temperature initial conditions, the full Thermo Field Dynamics formalism must now be employed. Within this framework, thermal states and observables can be consistently constructed, allowing their physical consequences to be investigated. This analysis is presented in the following section.

\section{Dual Thermo Field Dynamics in Bunch-Davies vaccum}\label{4}

In this section, finite-temperature effects are introduced into the global vacuum state $\ket{0_\text{BD}}\in \mathcal{H}_0=\mathcal{H}_R\otimes\mathcal{H}_L$. To this end, an auxiliary copy of the Hilbert space is introduced, together with the corresponding vacuum state $\ket{\widetilde{0}_\text{BD}}\in\widetilde{\mathcal{H}}_0=\widetilde{\mathcal{H}}_R\otimes\widetilde{\mathcal{H}}_L$. This doubling of degrees of freedom follows the standard Thermo Field Dynamics construction and provides the framework required to describe thermal states of the scalar field in the expanding universe. The formalism is analogous to that developed in Minkowski spacetime \cite{Khanna1,Khanna2}. The total thermal Hilbert space is then given by
\begin{eqnarray}
\mathcal{H}_\beta=\mathcal{H}_0\otimes\widetilde{\mathcal{H}}_0=\mathcal{H}_R\otimes\widetilde{\mathcal{H}}_R\otimes\mathcal{H}_L\otimes\widetilde{\mathcal{H}}_L,
\end{eqnarray}
which consists of four sectors associated with the right and left regions and their corresponding tilde counterparts. Within this framework, the field quartet is defined as
\begin{eqnarray}
	\Phi=\begin{pmatrix}
		\phi^R\\\phi^{L\;\dagger}\\\widetilde{\phi}^{R\;\dagger}\\\widetilde{\phi}^{L}
	\end{pmatrix},\quad\quad\text{and}\quad\quad \overline{\Phi}=\begin{pmatrix}
		\phi^{R\;\dagger}&-\phi^{L}&-\widetilde{\phi}^{R}&\widetilde{\phi}^{L\;\dagger}
	\end{pmatrix}.
\end{eqnarray}
The correspondence between tilde and non-tilde operators is established through the standard bosonic tilde-conjugation rules \cite{Book}. For an arbitrary operator $\mathcal{A}_i$, these relations take the form
\begin{eqnarray}
\widetilde{\mathcal{A}_i\mathcal{A}_j}=\widetilde{\mathcal{A}}_i\widetilde{\mathcal{A}}_j;\quad\quad\widetilde{\widetilde{\mathcal{A}}}=\mathcal{A};\quad\quad \widetilde{k\mathcal{A}_i+\mathcal{A}_j}=k^*\widetilde{\mathcal{A}}_i+\widetilde{\mathcal{A}}_j;\quad\quad \widetilde{\mathcal{A}^\dagger}=\left(\widetilde{\mathcal{A}}\right)^\dagger,
\end{eqnarray}
where $k$ is a complex constant. The commutation relations read
\begin{eqnarray}
	\left[\mathcal{A}_i,\mathcal{A}_j\right]=i\varepsilon^k_{ij}\mathcal{A}_k;\quad\quad \left[\widetilde{\mathcal{A}}_i,\widetilde{\mathcal{A}}_j\right]=-i\varepsilon^k_{ij}\widetilde{\mathcal{A}}_k;\quad\quad \left[\mathcal{A}_i,\widetilde{\mathcal{A}}_j\right]=0,
\end{eqnarray}
with $\varepsilon^k_{ij}$ being the structure constants. This algebraic structure justifies the doublet definition introduced in Eq.~\eqref{eq11}.

In the global Bunch-Davies frame, the non-thermal and thermal creation and annihilation operators are defined via
\begin{eqnarray}
	\mathcal{B}_{\ell m}^\alpha(\omega)=\begin{pmatrix}
		b^{\alpha}_{\ell m R}(\omega)\\b^{\alpha\dagger}_{\ell m L}(\omega)\\\widetilde{b}^{\;\alpha\dagger}_{\ell m R}(\omega)\\\widetilde{b}^{\;\alpha}_{\ell m L}(\omega)
	\end{pmatrix};\quad\quad  \mathcal{B}_{\ell m}^\alpha(\omega;\beta)=\begin{pmatrix}
		b^{\alpha}_{\ell m R}(\omega;\beta)\\b^{\alpha\dagger}_{\ell m L}(\omega;\beta)\\\widetilde{b}^{\;\alpha\dagger}_{\ell m R}(\omega;\beta)\\\widetilde{b}^{\;\alpha}_{\ell m L}(\omega;\beta)
	\end{pmatrix}.\label{eq15}
\end{eqnarray}
This construction allows the physical finite-temperature operators, $\mathcal{B}_{\ell m}^\alpha(\omega;\beta)$, to be expressed through the generalized Bogoliubov transformation $\mathcal{B}_{\ell m}^\alpha(\omega;\beta)=\mathbb{N}(\beta)\mathcal{B}_{\ell m}^\alpha(\omega)$, where
\begin{eqnarray}
	\mathbb{N}(\beta)=\begin{pmatrix}
		\cosh{\theta_\omega(\beta)} & 0 & \sinh{\theta_\omega(\beta)} & 0 \\
		0 & \cosh{\theta_\omega(\beta)} & 0 & \sinh{\theta_\omega(\beta)} \\
		\sinh{\theta_\omega(\beta)} & 0 & \cosh{\theta_\omega(\beta)} & 0 \\
		0 & \sinh{\theta_\omega(\beta)} & 0 & \cosh{\theta_\omega(\beta)} 
	\end{pmatrix}.\label{eq21}
\end{eqnarray}

The action of the operators $b^{\alpha\,\dagger}_{\ell mR}$ and $b^{\alpha}_{\ell mR}$ on the thermal Hilbert space is obtained by extending the zero-temperature relations of Eq.~\eqref{eq10} to finite temperature,
\begin{eqnarray}
    b^{\alpha}_{\ell mR}(\omega;\beta)\ket{0_\text{BD}(\beta)}=0,\quad\quad    \text{with} \quad\quad \left[b_{\ell m R}(\omega;\beta),b_{\ell^\prime m^\prime R}^\dagger(\omega^\prime;\beta)\right]=\delta(\omega-\omega^\prime)\delta_{\ell,\ell^\prime}\delta_{m,m^\prime}.
\end{eqnarray}
Similar relations hold for the operators in the left region. Within this framework, the thermal vacuum is constructed in the same manner as in Eq.~\eqref{eq17} and is given by
\begin{eqnarray}
    \ket{0_\text{BD}(\beta)}=\sqrt{1-e^{-\beta\omega}}\sum_{n=0}^{\infty}e^{-n\beta\omega/2}\ket{n_\text{BD}}\otimes\ket{\widetilde{n}_\text{BD}},
\end{eqnarray}
where $\ket{n_\text{BD}}=\ket{n^R_\text{BD}}\otimes\ket{n^L_\text{BD}}$ and $\ket{\widetilde{n}_\text{BD}}=\ket{\widetilde{n}^R_\text{BD}}\otimes\ket{\widetilde{n}^L_\text{BD}}$.

It is worth noting that an interesting feature emerges when the temporal evolution of the Bogoliubov angle is considered. Recalling Eq.~\eqref{eq19}, we obtain
\begin{eqnarray}
	\dot{\theta}_\omega=-\frac{1}{2}\sinh(\theta)\cosh(\theta)\frac{d\left(\beta\omega\right)}{dt}. 
\end{eqnarray}

\begin{figure}[ht]
	\centering
	
	\begin{subfigure}[ht]{0.48\textwidth}
		\centering
		\includegraphics[width=\linewidth]{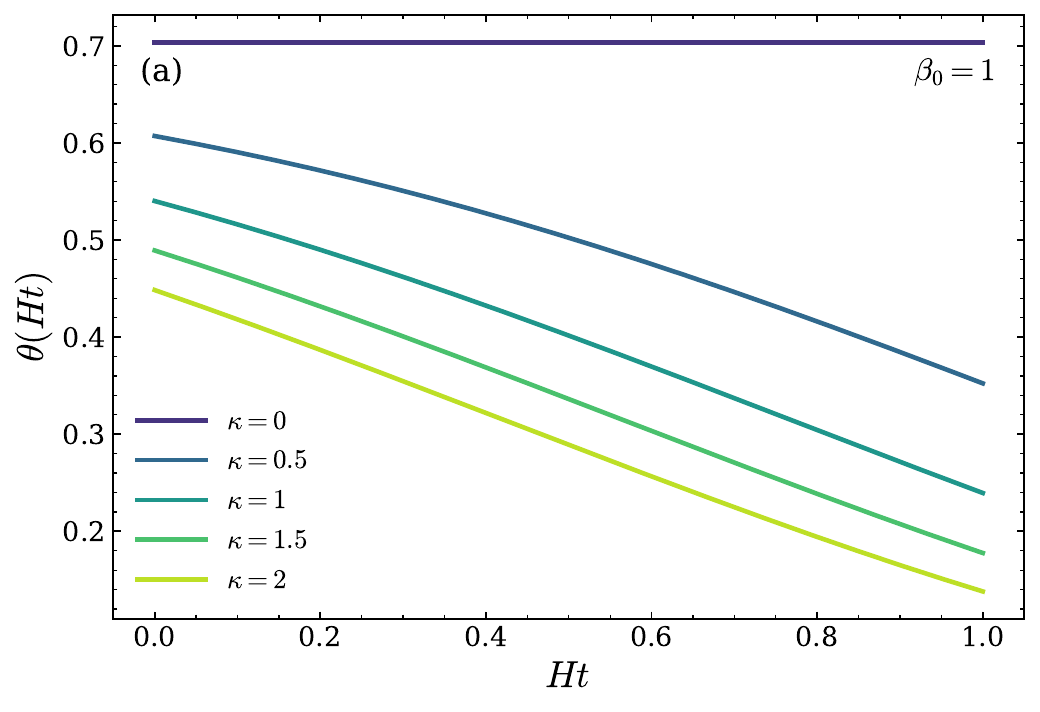}
	\end{subfigure}
	\hfill
	\begin{subfigure}[ht]{0.48\textwidth}
		\centering
		\includegraphics[width=\linewidth]{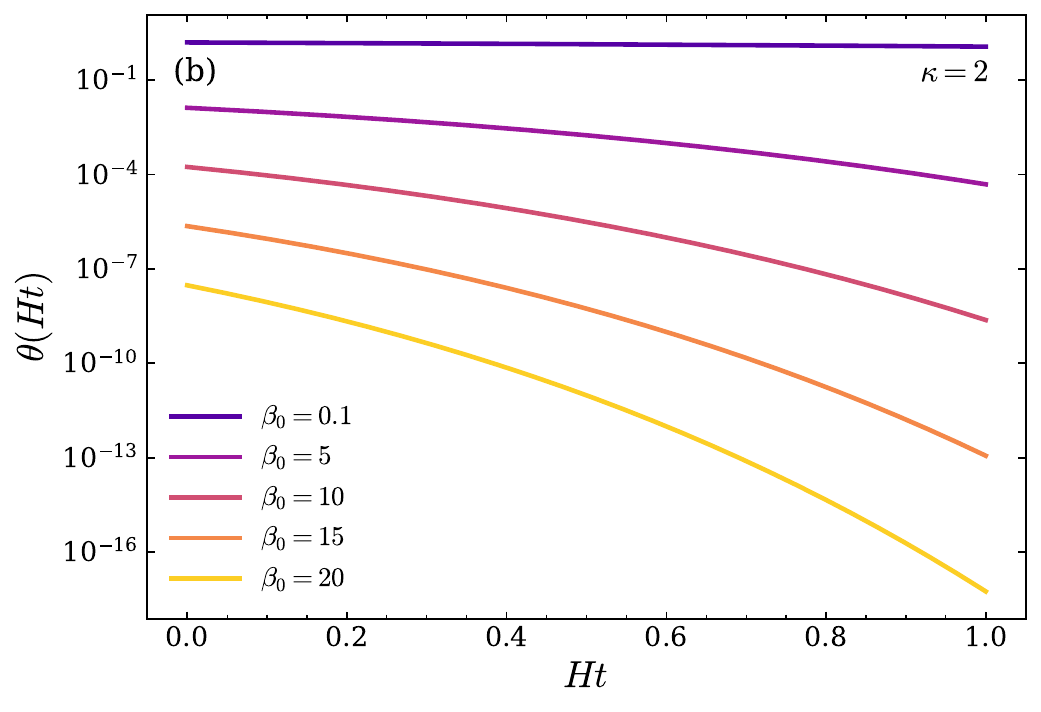}
	\end{subfigure}
	
	\caption{
		Evolution of $\theta(Ht)$ for different parameter choices.
		(a) Linear-scale evolution for fixed $\beta_0=1$ and varying $\kappa$.
		(b) Semilogarithmic evolution for fixed $\kappa=2$ and varying $\beta_0$. These plots are obtained by taking $p=1$.}
	
	\label{fig1}
\end{figure}

In the UV limit of the global solution describing the observable universe, as given in Eq.~\eqref{eq18}, it is found that $\dot{\theta}=0$ for modes with $p \to \infty$. This indicates that, in the Minkowski limit, the particle number density remains constant. Recalling Eq.~\eqref{eq16}, in this regime one has $\omega \approx p/a$, which leads to
\begin{eqnarray}
\beta = a(t)\,\beta_0,\label{eq27}
\end{eqnarray}
where $\beta_0$ is the initial inverse temperature, assumed to be constant throughout the cosmic evolution. This relation corresponds to Tolman's law \cite{tolman1930temperature} and must be satisfied by all modes for different values of $p$. Consequently, we can write
\begin{eqnarray}
	\dot{\theta}_\omega(t)=-\frac{1}{2}\left[\frac{H\beta_0\kappa\, e^{Ht}}{\sqrt{p^2e^{-2Ht}+\kappa}}\right]\sinh(\theta)\cosh(\theta),\label{eq20}
\end{eqnarray}
where $\kappa=M^2+12\xi H^2$. For a minimally coupled massless field ($\kappa=M=\xi=0$), the limit $\dot{\theta}=0$ is recovered. Given the initial condition
\begin{eqnarray}
	\sinh{\theta_\omega(0)}=\left[e^{\beta_0\sqrt{p^2+\kappa}}-1\right]^{-\frac{1}{2}},\label{eq22}
\end{eqnarray}
the differential equation \eqref{eq20} is solved numerically, as shown in Figure~\ref{fig1}.

As seen in Figure~\ref{fig1}(a), for a specific mode with momentum $p=1$ and a fixed initial temperature, the value of $\kappa$ is directly related to the dynamics of the universe. Initially, $\theta$ is larger, indicating significant thermal effects and a highly populated thermal bath. As time progresses and the universe expands, for $\kappa\neq0$, these effects are suppressed and tend toward zero. In this regime, the Bogoliubov transformation \eqref{eq21} becomes the identity, causing thermal observables to reduce to non-thermal quantities ($b(\omega;\beta)\to b(\omega)$), while the tilde space, associated with the thermal bath, becomes effectively inaccessible. The angle decreases more rapidly for higher values of $\kappa$, and the configuration of these parameters depends entirely on the field mass and the strength of the non-minimal coupling between the field and gravity. Conversely, when $\kappa=0$, as previously discussed, the field behaves as standard radiation, and the particle distribution in the thermal bath remains constant, being determined exclusively by the inverse initial temperature of the universe for that specific mode ($p=1$), as shown in Eq.~\eqref{eq22}.

On the other hand, considering the contribution of the initial temperature to the evolution of the $p=1$ mode with fixed $\kappa=2$, one observes a general reduction of the Bogoliubov angle over time, as shown in Figure~\ref{fig1}(b). This reduction is more pronounced for smaller initial values of $\beta_0$. Specifically, when the universe starts at a higher temperature ($\beta_0\to0$), the time required for the Bogoliubov angle to vanish, and consequently for the thermal bath to disappear, is considerably longer than in a universe that begins at a lower temperature.

\begin{figure*}[ht]
	\centering
	
	\begin{subfigure}[ht]{0.48\textwidth}
		\centering
		\includegraphics[width=\linewidth]{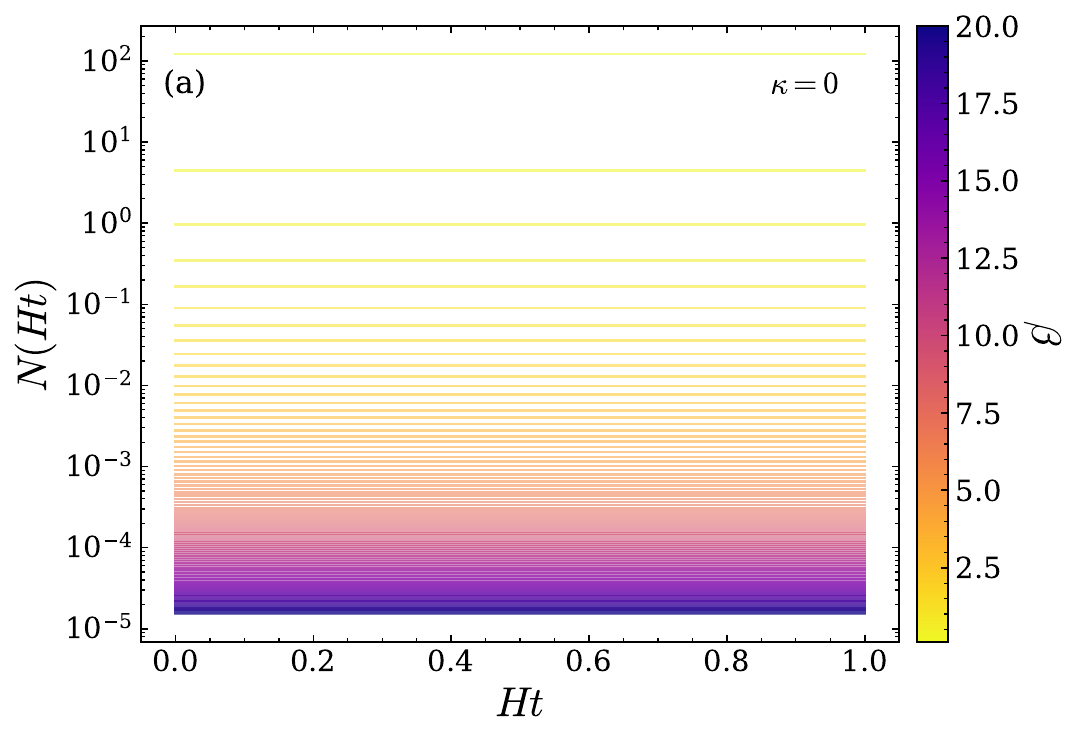}
		
	\end{subfigure}
	\hfill
	\begin{subfigure}[ht]{0.48\textwidth}
		\centering
		\includegraphics[width=\linewidth]{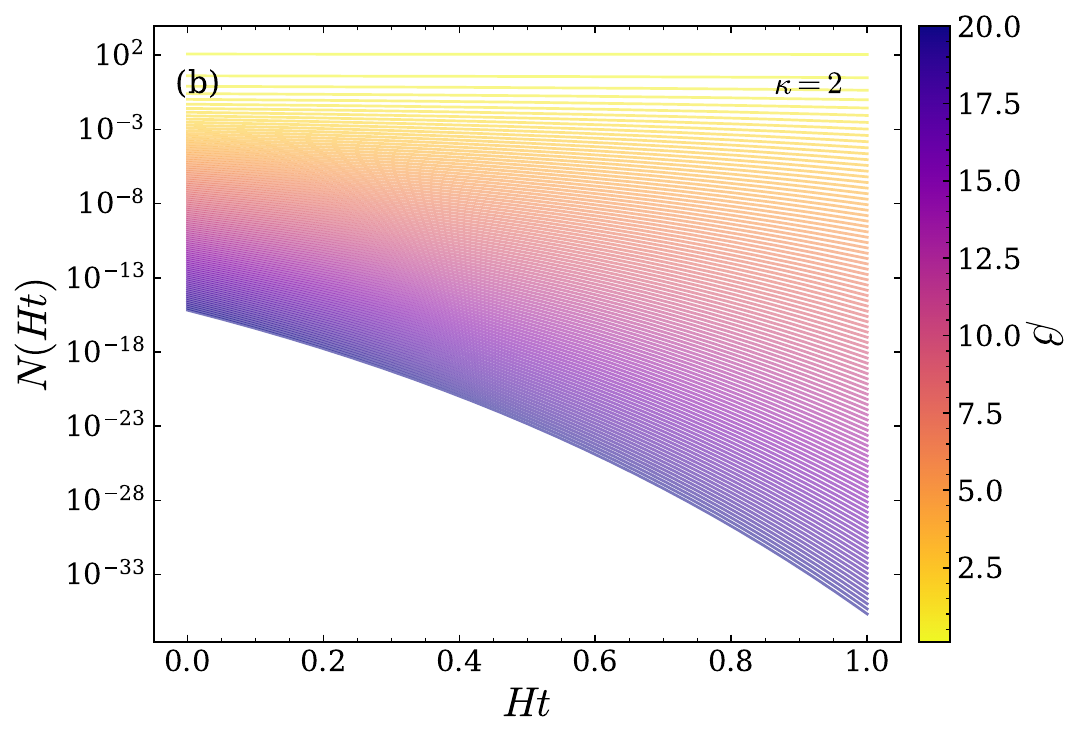}
		
	\end{subfigure}
	
	\vspace{0.4cm}
	
	\begin{subfigure}[ht]{0.48\textwidth}
		\centering
		\includegraphics[width=\linewidth]{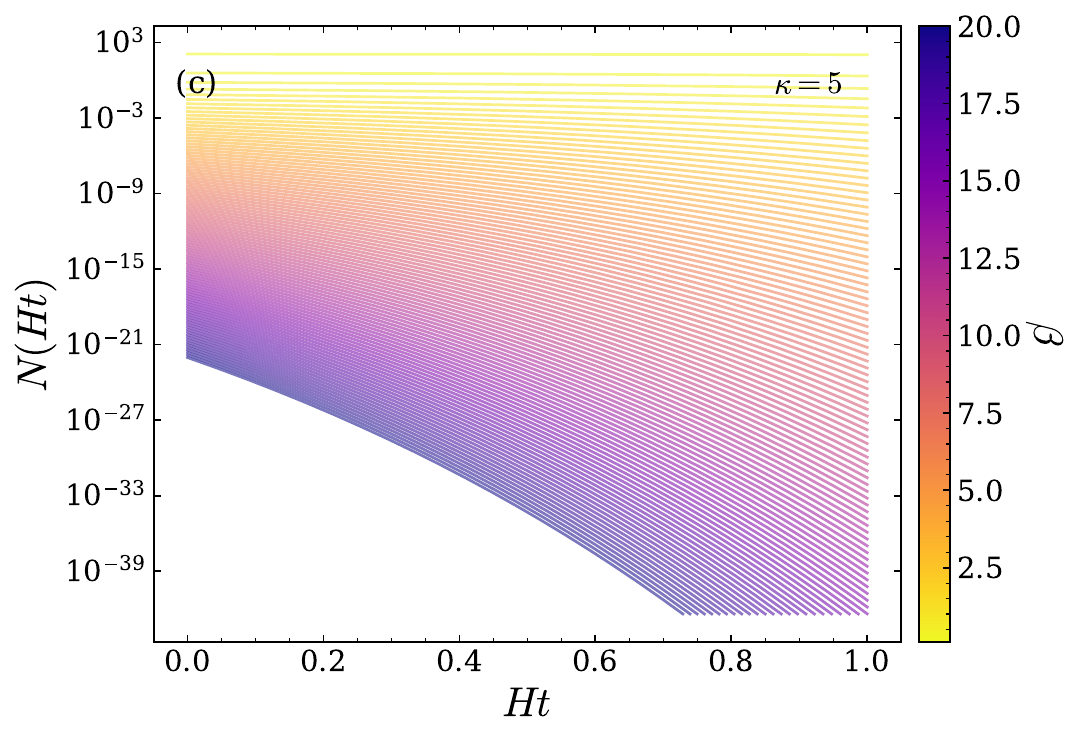}
		
	\end{subfigure}
	\hfill
	\begin{subfigure}[ht]{0.48\textwidth}
		\centering
		\includegraphics[width=\linewidth]{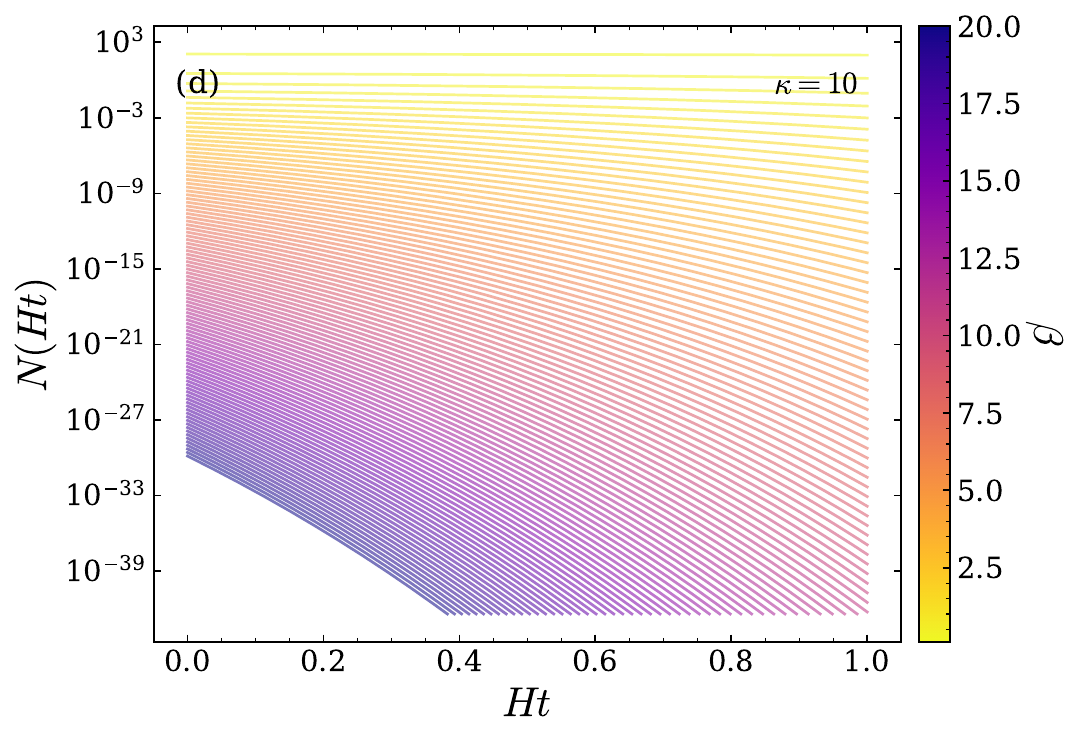}
		
	\end{subfigure}
	
	\caption{
		Evolution of the comoving number density $N(Ht)$ for different values of the inverse temperature parameter $\beta_0$ and fixed $\kappa$.
		Each curve corresponds to a different value of $\beta_0$, with the color scale indicating the thermal regime from high temperature (small $\beta_0$) to low temperature (large $\beta_0$).
		Panels correspond to:
		(a) $\kappa=0$,
		(b) $\kappa=2$,
		(c) $\kappa=5$,
		and (d) $\kappa=10$.
		The logarithmic scale highlights the different decay rates and asymptotic behaviors.
	}
	
	\label{fig2}
	
\end{figure*}

To account for the collective behavior of all field modes, rather than restricting the analysis to $p=1$, the comoving number density is evaluated, given by
\begin{eqnarray}
	N(t)=\frac{1}{2\pi^2}\int_0^\infty p^2\sinh^2{\theta_\omega(t)}dp,\label{eq24} 
\end{eqnarray}
defined as the number of particles per unit comoving volume, where $\theta_\omega(t)$ for each momentum $p$ is obtained by solving the differential equation \eqref{eq20}. The numerical evaluation of Eq.~\eqref{eq24} is presented in Figure~\ref{fig2}.

Consistent with the previous analysis, the parameter $\kappa$ exerts a significant influence not only on individual modes but also on the collective behavior of all modes. Assuming a nonzero $\kappa$ at the beginning of the universe, the comoving number density of the thermal plasma decreases considerably as the system evolves, similarly to the behavior observed for a single mode. The larger the magnitude of $\kappa$, the faster the thermal bath as a whole disappears. However, if the universe starts in a hotter state, this decay is delayed compared to universes that begin in colder states, as can be observed in Figures~\ref{fig2}(b), \ref{fig2}(c), and \ref{fig2}(d). On the other hand, in the radiation limit ($\kappa=0$), as previously discussed, the comoving number density remains constant throughout the cosmological expansion. This constant population is significantly larger for universes that begin at high initial temperatures (small $\beta_0$), while it remains much smaller for colder initial configurations, as shown in Figure~\ref{fig2}(a).

This specific behavior of the finite-temperature Bunch--Davies vacuum in the radiation-dominated case accurately reflects the dynamics of the Cosmic Microwave Background (CMB) \cite{starobinsky1982dynamics}. In particular, an initially very hot, radiation-dominated universe ($\kappa=0$) exhibits no change in its comoving number density. However, as the universe expands, the physical volume occupied by the thermal bath increases, and the typical separation between its constituents grows. This expansion reduces the interaction rate between particles and consequently lowers the temperature, in accordance with Tolman's law \eqref{eq27}. Ultimately, the universe cools to the point where the temperature associated with the scalar radiation becomes negligible, leading asymptotically to the standard zero-temperature FLRW cosmological regime.

In the static frame, the geometric Bogoliubov transformation given in Eq.~\eqref{eq14} maps the $b_{\ell m}$ operators into the $c_{\ell m}$ operators. Its generalization to four dimensions reads
\begin{eqnarray}
	\mathbb{M}(\beta_H)=\begin{pmatrix}
		\cosh{\theta_\omega(\beta_H)} &\sinh{\theta_\omega(\beta_H)} & 0 & 0\\
		\sinh{\theta_\omega(\beta_H)} & \cosh{\theta_\omega(\beta_H)} & 0 & 0\\
		0 & 0 & \cosh{\theta_\omega(\beta_H)} & \sinh{\theta_\omega(\beta_H)}\\
		0 & 0 & \sinh{\theta_\omega(\beta_H)} & \cosh{\theta_\omega(\beta_H)}
	\end{pmatrix},
\end{eqnarray}
such that the thermal operators observed in the static frame satisfy
\begin{eqnarray}
	\mathcal{C}_{\ell m}(\omega;\beta)=\mathbb{N}(\beta)\mathbb{M}(\beta_H)\mathcal{B}_{\ell m}(\omega),
\end{eqnarray}
where the matrix form $\mathcal{C}_{\ell m}(\omega;\beta)$ is obtained from Eq.~\eqref{eq15} by interchanging $b_{\ell m}\leftrightarrow c_{\ell m}$.

\begin{figure*}[ht]
	\centering
	\includegraphics[width=0.5\linewidth]{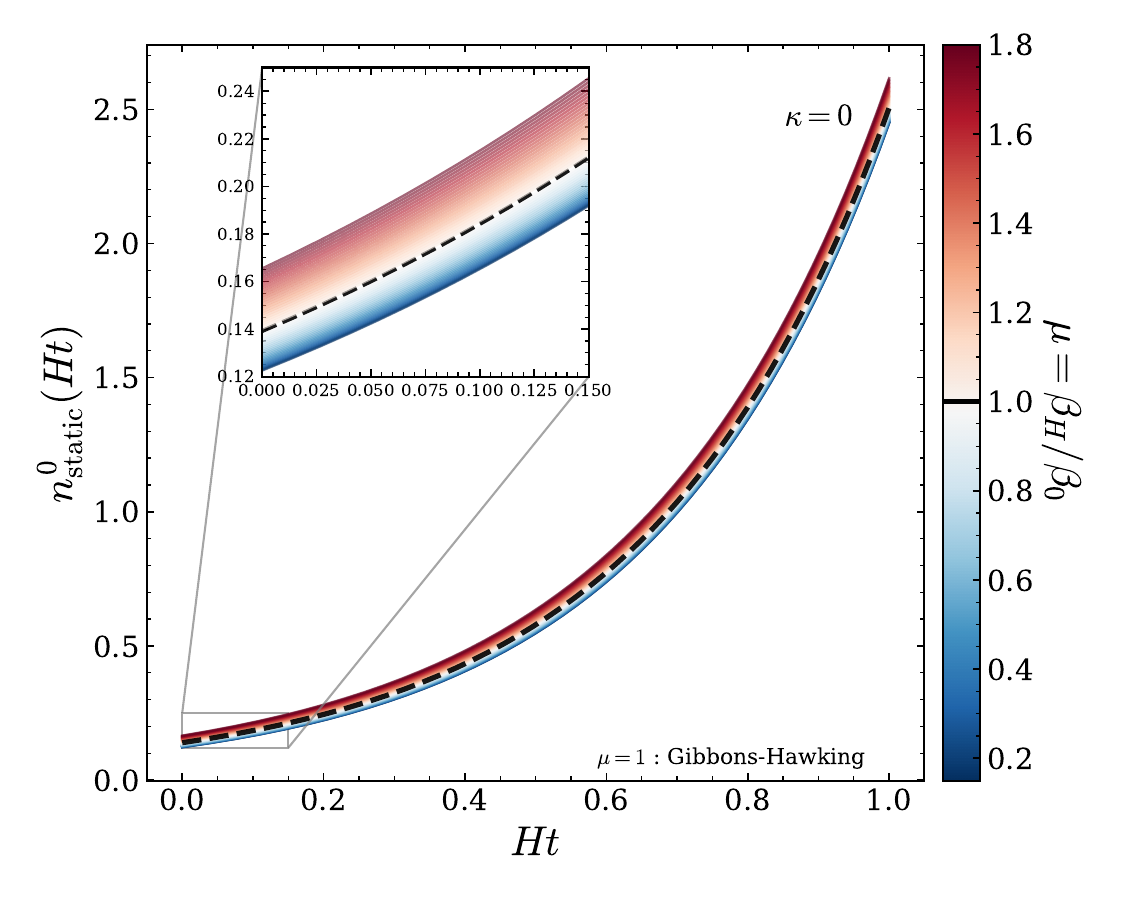}

	\caption{
		Evolution of the static-frame number density $n_{\mathrm{static}}(Ht)$ as a function of the dimensionless cosmological time $Ht$ for different values of the ratio $\mu=\beta_H/\beta_0$, where $\beta_H$ is the Gibbons-Hawking inverse temperature and $\beta_0$ is the inverse temperature associated with the static vacuum distribution at the beginning of the universe. The main graph corresponds to $\kappa=0$. The dashed black curve denotes the distinguished thermal configuration $\mu=1$, corresponding to the initial equilibrium at the Gibbons-Hawking temperature.}
	
	\label{fig3}
	
\end{figure*}

This construction enables the determination of finite-temperature quantities in the static frame in terms of the Bunch-Davies operators. Generalizing the result presented in Eq.~\eqref{eq25} to a global thermal vacuum with temperature $\beta^{-1}$, we obtain
\begin{eqnarray}
	\bra{0_\text{BD}(\beta)}c_{\ell mR}^{\alpha\;\dagger}(\omega;\beta)c_{\ell mR}^{\alpha}(\omega;\beta)\ket{0_\text{BD}(\beta)}=\frac{1}{e^{\beta_H\omega}-1}\left(1+\frac{1}{e^{\beta\omega}-1}\right),
\end{eqnarray}
which reveals an additional contribution to the geometric thermal bath perceived in the static frame. In analogy with Eq.~\eqref{eq24}, the number density measured by the static observer is given by
\begin{eqnarray}
	n_\text{static}(t)=\frac{1}{2\pi^2}\int_0^\infty \frac{p^2dp}{e^{\beta_H\omega}-1}\left[1+\sinh^2{\theta_\omega(t)}\right].\label{eq26}
\end{eqnarray}
The numerical evaluation of this quantity is shown in Figure~\ref{fig3}.

Considering first the radiation field ($\kappa=0$), the quantity in Eq.~\eqref{eq26} reduces to
\begin{eqnarray}
	n^0_\text{static}(t)=\frac{1}{2\pi^2}\int_0^\infty\frac{p^2dp}{e^{\beta_Hp/a(t)}-1}\left[1+\frac{1}{e^{\beta_0p}-1}\right].
\end{eqnarray}
In the limit $t\to\infty$, this quantity grows exponentially as $e^{Ht}$, describing an amplified particle emission phenomenon. Conversely, if the initial temperature of the universe is negligible, the dynamics associated with the Parker effect reduce to those described by Eq.~\eqref{eq25}, as expected.

On the other hand, for a nonzero initial temperature, one readily observes a significant enhancement in the particle population driven exclusively by the initial thermal bath. This behavior characterizes a stimulated emission process induced by the presence of the primordial thermal reservoir of the universe.

Figure~\ref{fig3} condenses the previous discussion in terms of a new dimensionless parameter $\mu=\beta_H/\beta_0$, illustrating the exponential growth for all configurations through a temperature heat map. Hotter universes ($\mu>1$) exhibit more pronounced stimulated Parker emission than colder universes ($\mu<1$), asymptotically approaching the zero-temperature effect described by Eq.~\eqref{eq25}. An interesting feature arises when $\beta_0=\beta_H$ ($\mu=1$): the universe initially remains in thermal equilibrium between the two reservoirs at the Gibbons-Hawking temperature, acting as an intermediate emission regime separating the hot and cold initial conditions of the universe.

\begin{figure}[ht]
	\centering
	
	\begin{subfigure}[ht]{0.48\textwidth}
		\centering
		\includegraphics[width=\linewidth]{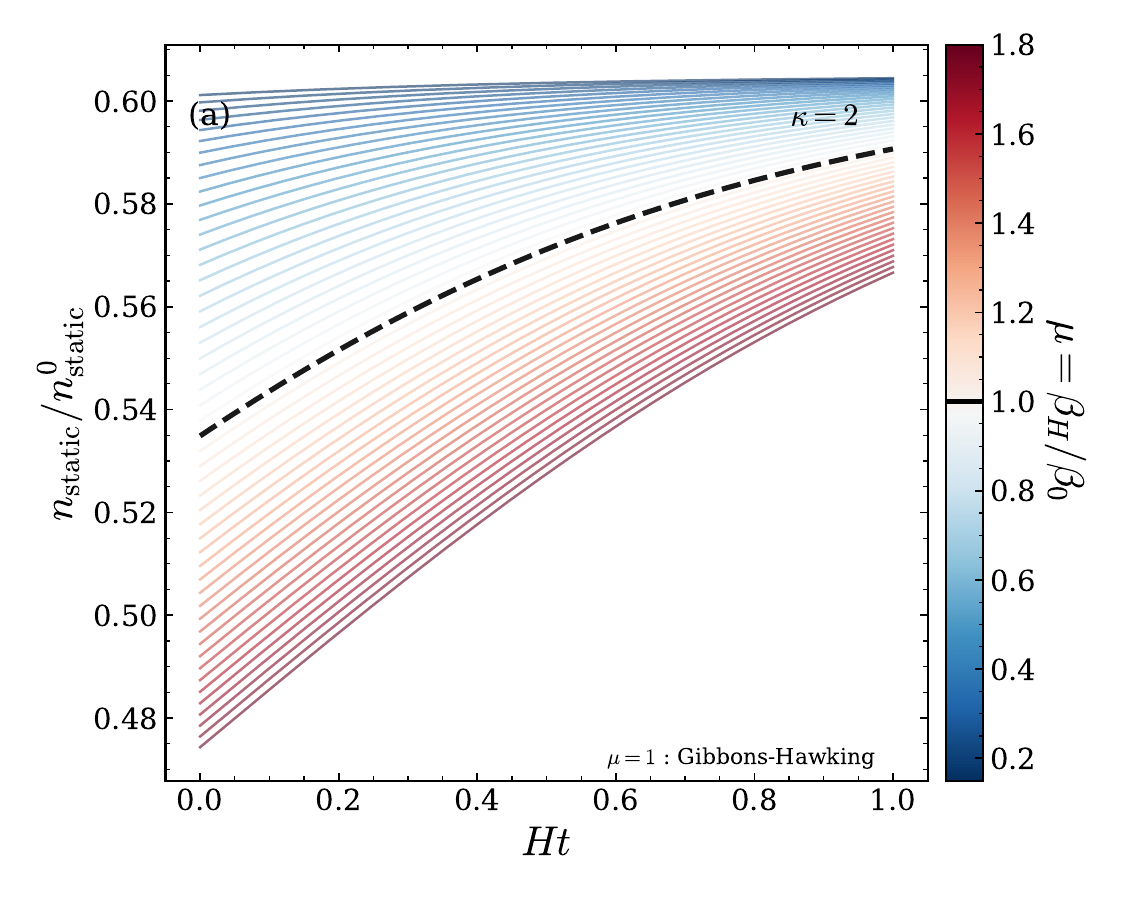}
	\end{subfigure}
	\hfill
	\begin{subfigure}[ht]{0.48\textwidth}
		\centering
		\includegraphics[width=\linewidth]{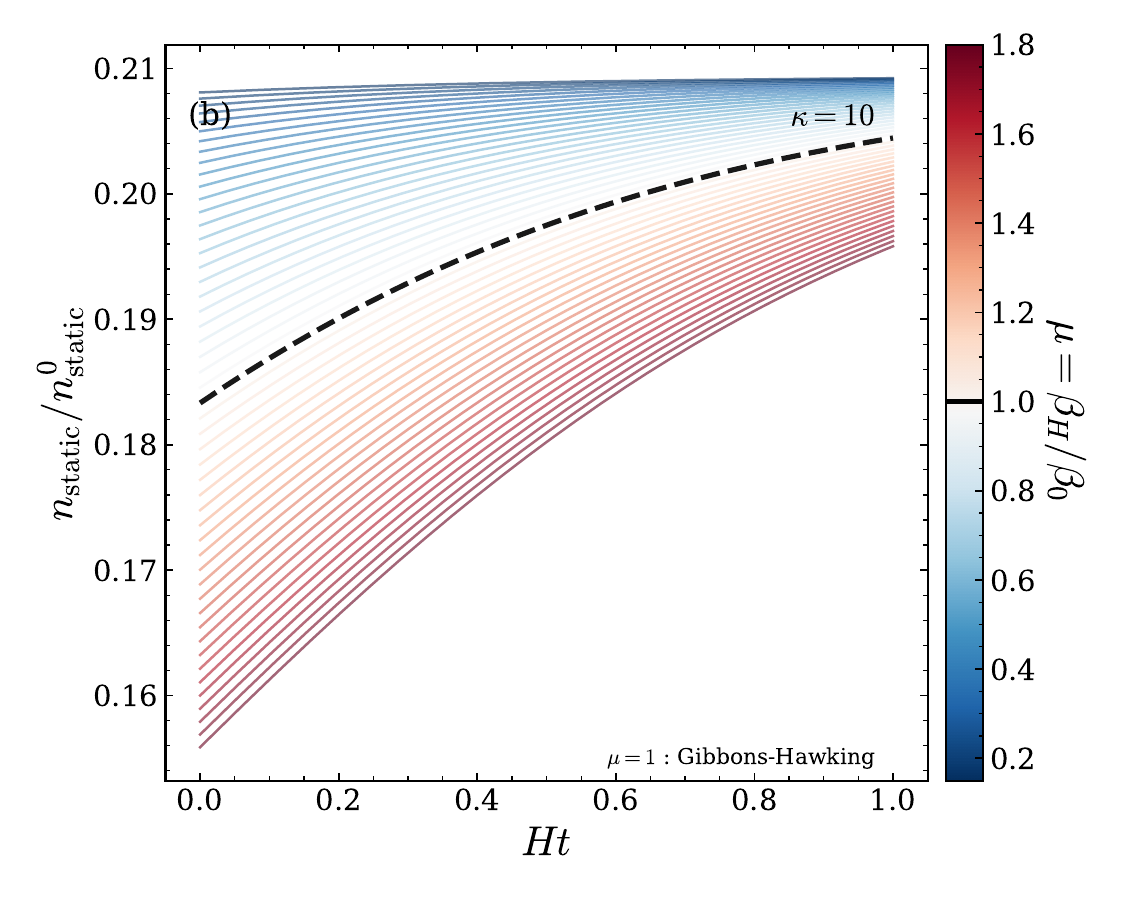}
	\end{subfigure}
	
	\caption{
		Evolution of the static-frame number density ratio $n_\text{static}/n^0_\text{static}$ between a $\kappa\neq0$ distribution and the radiation distribution as a function of the dimensionless cosmological time $Ht$ for different values of the ratio $\mu=\beta_H/\beta_0$, where $\beta_H$ is the Gibbons-Hawking inverse temperature and $\beta_0$ is the inverse temperature associated with the static vacuum distribution at the beginning of the universe. Panels correspond to:
		(a) $\kappa=2$ and (b) $\kappa=10$. The dashed black curve denotes the distinguished thermal configuration $\mu=1$, corresponding to the initial equilibrium at the Gibbons-Hawking temperature. 
	}
	
	\label{fig4}
\end{figure}

When the parameter $\kappa$ is taken into account, the evolution of the number density in the static frame, given by Eq.~\eqref{eq26}, differs significantly from the radiation case ($\kappa=0$), as shown in Figure~\ref{fig4}. By analyzing the ratio between the two cases, one readily observes a similar asymptotic behavior at late times, differing only in magnitude. Specifically, larger values of $\kappa$ (i.e., larger mass or stronger non-minimal coupling) suppress the time evolution of $n_\text{static}(t)$, leading to a delayed growth of the Parker emission effect in the limit $\kappa\gg1$ at fixed temperature.

On the other hand, examining the thermal contribution for a fixed non-zero $\kappa$, we observe an asymptotic regime in which the number density of a massive, non-minimally coupled field approaches the same exponential behavior as the radiation case, up to a multiplicative factor. This leads to a constant ratio at late times, a regime that is approached more slowly in hotter universes ($\mu>1$). For colder universes ($\mu<1$), the ratio reaches its asymptotic behavior much more rapidly, indicating that the detectability of the Parker emission magnitude for $\kappa>0$ fields strongly depends on the cosmological epoch under consideration. Once again, $\mu=1$ represents the initial thermal equilibrium of the universe at the Gibbons-Hawking temperature. Finally, the magnitude of this ratio decreases as $\kappa$ increases; consequently, the radiation case is significantly easier to detect than a massive, non-minimally coupled field within this thermal framework.

\section{Conclusions}\label{5}

In this work, the dynamics of a massive scalar field non-minimally coupled to gravity in an expanding de Sitter universe have been investigated. Both geometric and intrinsic thermal effects were incorporated within the framework of TFD through an appropriate doubling of the Hilbert space. The geometric temperature arises naturally as a consequence of the horizon structure and the change of reference frame used to describe the same quantum field. This mechanism leads to a geometric doubling of the quantum space, showing that the doubling procedure is not merely a mathematical artifact, but rather a manifestation of the global causal structure of spacetime. In contrast, the intrinsic temperature originates from the presence of a genuine thermal bath, which maintains the quantum field in a finite-temperature state and gives rise to statistical thermal effects beyond those induced by the horizon itself.

A complete thermal formulation of the field theory was developed, allowing both the intrinsic and purely geometric properties of the quantum vacua to be analyzed. The distinction between these two notions of temperature proves to be essential for understanding the thermal evolution of the universe. The temporal dynamics of the thermal bath were determined, and the particle number densities were evaluated in both comoving and static frames. In the comoving frame, the particle number density remains constant in the radiation limit, whereas for a massive non-minimally coupled field it decreases as the universe expands. Consequently, an initially hot, radiation-dominated universe embedded in a thermal plasma undergoes a continuous cooling process, accompanied by the dilution of the thermal bath. In the radiation regime, the conservation of the comoving particle density provides a thermodynamic evolution consistent with that expected for the Cosmic Microwave Background.

In the static frame, the standard zero-temperature Parker effect is recovered. At finite temperature, this mechanism manifests itself through the stimulated emission of scalar particles, which, in the radiation limit, exhibits an exponential growth with the dimensionless cosmological time $Ht$. In contrast, for massive or non-minimally coupled fields, the dynamics depend sensitively on both the initial temperature of the universe and the magnitude of the field parameters. In this regime, the intrinsic thermal bath enhances the geometric particle creation process. A particularly interesting behavior emerges when the universe is initially in thermal equilibrium at the Gibbons--Hawking temperature, signaling the existence of a characteristic temperature scale governing the interplay between geometric and statistical thermal effects. Finally, the detectability of Parker emission associated with highly massive and strongly coupled particles is found to depend strongly on the cosmological epoch at which the observation is performed.

\section*{Acknowledgments}

This work by A. F. S. is partially supported by National Council for Scientific and Technological
Development - CNPq project No. 312406/2023-1. D. S. C., L. A. S. E. and J. C. R. S. thank CAPES for financial support. This work was supported by computational resources provided by the Centro Nacional de Processamento de Alto Desempenho em São Paulo (CENAPAD--SP).

\section*{Data Availability Statement}

No Data associated in the manuscript.

\section*{Conflicts of Interest}

No conflict of interests in this paper.


\global\long\def\link#1#2{\href{http://eudml.org/#1}{#2}}
\global\long\def\doi#1#2{\href{http://dx.doi.org/#1}{#2}}
\global\long\def\arXiv#1#2{\href{http://arxiv.org/abs/#1}{arXiv:#1 [#2]}}
\global\long\def\arXivOld#1{\href{http://arxiv.org/abs/#1}{arXiv:#1}}


\end{document}